\documentclass[%
 prb,
 amsmath,amssymb,
 reprint,%
 superscriptaddress,
]{revtex4-1}

\usepackage{graphicx}
\usepackage{dcolumn}
\usepackage{bm}

\usepackage[utf8]{inputenc}
\usepackage[T1]{fontenc}
\usepackage{mathptmx}

\begin{document}

\title[Multi-channel scattering mechanism behind the re-entrant conductance feature in nanowires subject to strong spin-orbit coupling]
  {Multi-channel scattering mechanism behind the re-entrant conductance feature in nanowires subject to strong spin-orbit coupling}

\author{Iann Cunha}
\affiliation{Departamento de F\'{\i}sica, Universidade Federal de S\~ao Carlos, 13565-905, S\~ao Carlos, SP, Brazil}
\author{Leonardo Villegas-Lelovsky}
\affiliation{Departamento de F\'{\i}sica, Universidade Federal de S\~ao Carlos, 13565-905, S\~ao Carlos, SP, Brazil}
\affiliation{Departamento de F\'{\i}sica, IGCE, Universidade Estadual Paulista, 13506-900 Rio Claro SP, Brazil}
\author{Victor Lopez-Richard}
\affiliation{Departamento de F\'{\i}sica, Universidade Federal de S\~ao Carlos, 13565-905, S\~ao Carlos, SP, Brazil}
\author{Leonardo Kleber Castelano}
\email{lkcastelano@ufscar.br}
\affiliation{Departamento de F\'{\i}sica, Universidade Federal de S\~ao Carlos, 13565-905, S\~ao Carlos, SP, Brazil}

\date{\today}

\begin{abstract}
The characterization of helical states can be performed by checking the existence of the re-entrant behaviour, which appears as a dip in the conductance probed in nanowires (NWs) with strong spin-orbit coupling (SOC) and under perpendicular magnetic field. On the other hand, the experiment described in Ref.~\citenum{Heedt} observed the re-entrant behaviour in the absence of a magnetic field, which was explained through spin-flipping two-particle backscattering. We theoretically show that the observation of the re-entrant behaviour is due to a multi-channel scattering mechanism, which causes a reduction of the transmission when an effective attractive potential and coupling between different channels are present. Both ingredients are provided by the SOC in the transport properties of NWs.
\end{abstract}

\maketitle

 Topologically protected quantum computation can be achieved by employing Majorana zero modes~\cite{Nayak}. Such Majorana states were predicted to be observed in nanowires (NWs) with strong spin-orbit coupling (SOC) in proximity to a s-wave superconductor and under the presence of an external magnetic field~\cite{Alicea,Oreg,Lutchyn}. Semiconductor NWs based on InAs and InSb have a strong SOC and the experimental realization and characterization of such NWs have been recently explored to check the existence of helical states~\cite{Heedt,Kammhuber,Weperen,Sun}, which are closely related to Majorana zero modes. A signature to verify helical states is the called re-entrant behaviour~\cite{Streda}, which appears as a measurable dip in the conductance when an external magnetic field is strictly different from zero. In Ref.~\citenum{Heedt}, an unexpected result was observed, which was the appearance of the re-entrant characteristic in the absence of the external magnetic field. This feature was attributed to spin-flipping two-particle backscattering~\cite{Heedt,Khrapai}. Nonetheless, the dip at zero magnetic field also happens when there is scattering by impurities~\cite{Saldana,Bardarson}. Consequently, a clear explanation for the re-entrant feature is still lacking~\cite{Sun} and we also subscribe that assertion.
 
 In this Letter, we want to shed light on the problem of the observation of the re-entrant behaviour when the magnetic field is absent by adding an essential ingredient that has clearly been previously overlooked. The preliminary theoretical prediction of a re-entrant behaviour of the conductance was ascribed to the opening of a pseudogap in the energy dispersion of the NW by combining SOC and a magnetic field~\cite{Streda}. Both theoretical and experimental validations of such a prediction in InAs and InSb NWs have been attempted for some time~\cite{Kammhuber,Weperen,Heedt,Pershin,Tang,Rainis}. However, the result found in Ref.~\citenum{Heedt} is surprising because there is no pseudogap when the magnetic field is absent. In turn, theoretical models that predict the pseudogap opening consider the concurrence of three factors: the SOC along the whole NW~\cite{Streda,Tang,Rainis}, a magnetic field, and a spatial potential localized in a finite region, which controls the re-entrant behavior by tuning the energy into the pseudogap. Particularly, in the experiment described in Ref.~\citenum{Heedt}, the perpendicular electric field is applied in a finite region  along the longitudinal direction of the NW as depicted in Fig.~1. Once an electric field is applied, the induced structural inversion asymmetry SOC emerges. Thus, the bias voltage applied between source and drain as depicted in Fig.~1 also triggers a SOC, but the electric field $E_x$ applied between top- and  bottom-gate is much higher; thereby, causing a strong SOC in the finite region of length $L$. In this letter we will unveil how this last external field configuration plays an important role in the transport properties and provide a clear explanation for the re-entrant behaviour in the absence of the magnetic field. To understand this effect, one needs to know that the Rashba Hamiltonian limited to a finite region behaves as an attractive potential, as shown in Ref.~\citenum{Sanchez}. Furthermore, it has been demonstrated that a resonant reflection (dip in the conductance) might occur if two conditions are satisfied: the scattering by an attractive potential and the coupling between different transport channels~\cite{Gurvitz}, even in the absence of a magnetic field. Precisely, these two ingredients are present in quasi-1D systems, such as NWs, where the SOC takes place in a finite region. 
\begin{figure}[tb]
\includegraphics[scale=.3,angle=00]{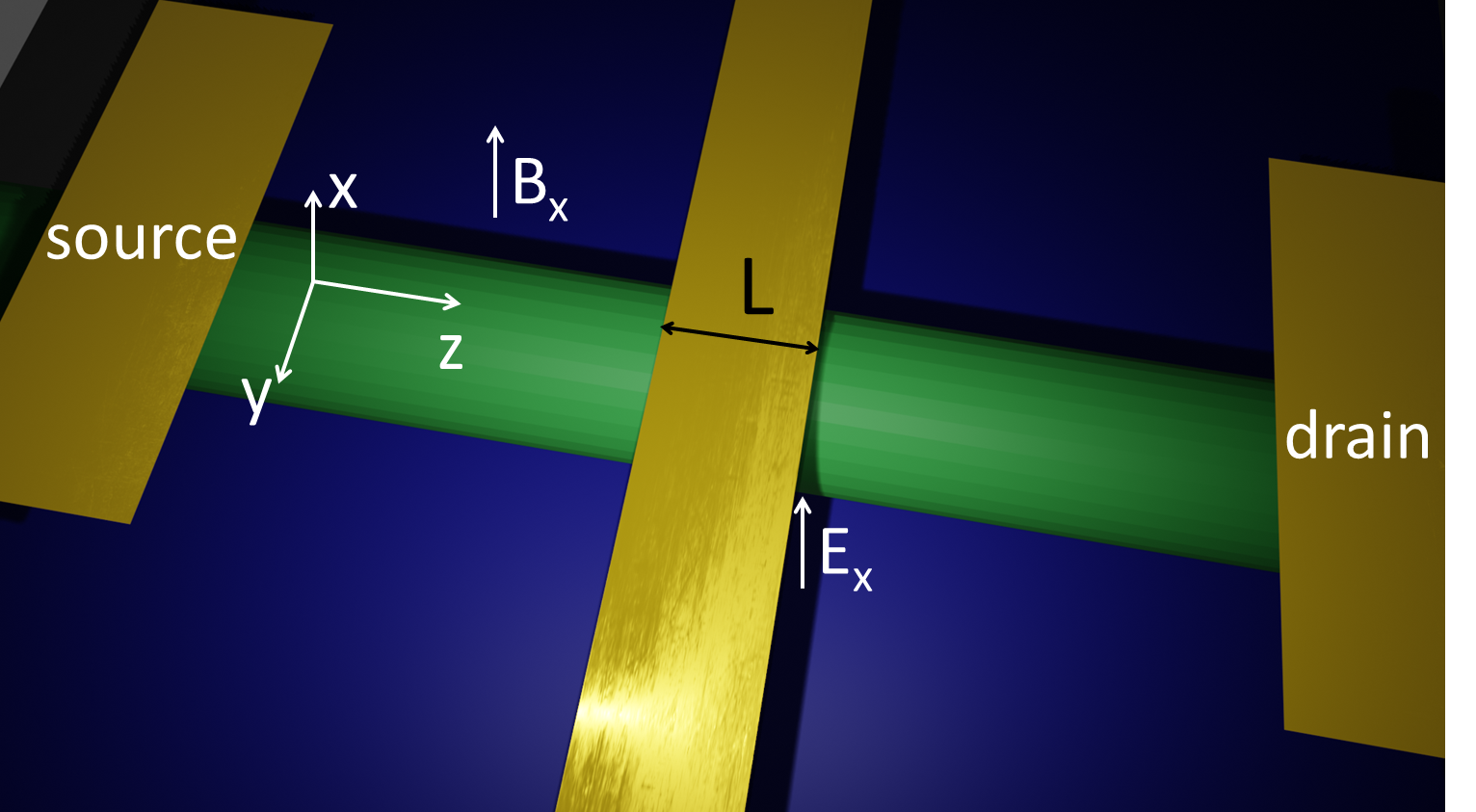}\caption{Schematic picture showing the NW, whose longitudinal direction coincides with the $z-$direction, the source and the drain, the region with length $L$ where the perpendicular electric field $E_x$ between top- and bottom-gate is applied, and the magnetic field $B_x$ applied in the $x-$direction.}%
\end{figure}

We theoretically simulate the electronic transport through the NW where the Rashba SOC occurs in a region with length $L$ along the z-direction according to the following Hamiltonian
 \begin{equation}
H=\left(\frac{p^2}{2m}+V_c(\vec{r})\right){\bf I}+H_R. \label{eq1}
\end{equation}
The first term is the kinetic energy, $V_c(\vec{r})$ is the confining potential, $H_R$ is the Rashba Hamiltonian, and $\bf{I}$ is the 2$\times$2 identity matrix. The confining potential represents the lateral confinement of the NW, which is modelled as $V_c(\rho)=0$ for $\rho\leq r_0$ and $V_c(\rho)=\infty$ for $\rho> r_0$, where $\rho=\sqrt{x^2+y^2}$ and $r_0$ is its radius. Particularly, we assume that the Rashba Hamiltonian occurs only in the region of length $L$, thus $H_R=\alpha f(z)(k_y\sigma_z-k_z\sigma_y)$, where $f(z)=\Theta(z+L/2)-\Theta(z-L/2)$ and  $\Theta(z)$ is the Heaviside function, $\alpha$ is the Rashba coupling constant, $k_q$ and $\sigma_q$ are the wave-vector and the Pauli spin matrix in the $q-$direction, respectively.
 The wave-function $\Psi(\vec{r})$ can be expanded as a function of the lateral eigenfunctions $\phi_{n,m}(\rho,\varphi)=A_{n,m}J_n(\mu_{n,m}\rho/r_0)e^{in\varphi}$, which are solutions of the two-dimensional hard wall problem whose eigenenergies are $\epsilon_{n,m}=\hbar^2\mu_{n,m}^2/2mr_0^2$, where $n=0,\pm 1,\pm 2,\ldots$ and $m=1,2,\ldots$. The normalization constant is $A_{n,m}=1/[r_0\sqrt{\pi}J_{n+1}(\mu_{n,m})$ and the \textit{m}th zero of the \textit{n}th-order Bessel function is denoted by $\mu_{n,m}$. The wave-function has two spin components $\Psi^{\pm}(\vec{r})=\sum_{n}\sum_{m} \psi^{\pm}_{n,m}(z)\phi_{n,m}(\rho,\varphi)$, where $\phi^{s}_{n,m}(z)$ is the scattering wave-function in the z-direction for each transport channel $j=(n,m,s)$, where $s=\pm$. By employing this expansion and the Eq.(\ref{eq1}), we can find the following matrix elements
\begin{equation}
H_{j,j\prime}=\left[\frac{\hbar^2k_z^2}{2m}+\epsilon_j\right]~\delta_{j,j\prime}+ c_{j,j\prime}+d_{j,j\prime}.\label{eq2}
\end{equation}
In Eq.~(\ref{eq2}), we have the spin degenerated orbital eigenenergy $\epsilon_j=\epsilon_{n,m}$, the intersubband term $c_{j,j\prime}=\alpha(z)\left\langle j|k_y\sigma_z|j\prime\right\rangle$, and the intrasubband term $d_{j,j\prime}=-\left\langle j|\{\alpha(z),k_z\}\sigma_y|j\prime\right\rangle$, where $\{\alpha(z),k_z\}=(\alpha(z)k_z+k_z\alpha(z) )/2$ and $\alpha(z)=\alpha f(z)$.  These terms can be analytically calculated as follows~\cite{leo}: $c_{j,j\prime}=s\prime\delta_{s,s\prime}\alpha(z)T^{n,m}_{n\prime,m\prime}(\delta_{n\prime,n+1}-\delta_{n\prime,n-1})/r_0$, where $T_{n,m}^{n\prime,m\prime}=\mu_{n,m}/\mu_{n\prime,m\prime}/[(\mu_{n,m}/\mu_{n\prime,m\prime})^2-1]$ and $d_{j,j\prime}=-is\prime\{\alpha(z),k_z\}\delta_{n,n\prime}\delta_{m,m\prime}\delta_{s,-s\prime}$. Ideally, we should consider infinite transport channels (subbands) to completely solve the transport along the z-direction, but the influence of high energy channels would exponentially decay if incoherent transport effects were considered.~\cite{winkler}

To extract the basic transport properties of this problem, we assume four coupled channels with lower energies. Within this subspace, we get the following expression for the Hamiltonian

\begin{eqnarray}
H=\left( \begin{array}{cccc} \frac{\hbar^2k_z^2}{2m}+\epsilon_1 & i\{{\alpha},~k_z\}& {\alpha}\frac{T^{0,1}_{1,1}}{r_0}&0 \\ -i\{{\alpha},~k_z\} & \frac{\hbar^2k_z^2}{2m}+\epsilon_2&0&-{\alpha}\frac{T^{0,1}_{1,1}}{r_0}\\ -{\alpha}\frac{T^{1,1}_{0,1}}{r_0}&0& \frac{\hbar^2k_z^2}{2m}+\epsilon_3&i\{{\alpha},~k_z\}\\ 0&{\alpha}\frac{T^{1,1}_{0,1}}{r_0}&-i\{{\alpha},~k_z\}& \frac{\hbar^2k_z^2}{2m}+\epsilon_4  \end{array} \right),\label{eq3}
\end{eqnarray}
where $j=1\Rightarrow(0,1,+)$, $j=2\Rightarrow(0,1,-)$,  $j=3\Rightarrow(1,1,+)$, and $j=4\Rightarrow(1,1,-)$. In this case, we have that $\epsilon_1=\epsilon_2=\epsilon$, $\epsilon_3=\epsilon_4=\lambda$, $r(z)={\alpha}(z)T^{0,1}_{1,1}/r_0=-{\alpha}(z)T^{1,1}_{0,1}/r_0$,  and the transport wave function is $\Phi=\left(\psi^+_{0,1}(z)\; \psi^-_{0,1}(z)\; \psi^+_{1,1}(z)\; \psi^-_{1,1}(z)\right)^t$. By introducing the following linear combination between transport channels with same energy $\chi^\pm_{n,m}(z)=(\psi^+_{n,m}(z)\pm i\psi^-_{n,m}(z))e^{\mp i\theta(z)}/\sqrt{2}$, where $k_\alpha(z)=m{\alpha}(z)/\hbar^2$ and $e^{\pm i\theta(z)}=\exp\left(\pm i\int_0^zk_\alpha(z\prime)dz\prime\right)$ we can write the transport wave function in this new basis $\Phi=\left(\chi^+_{0,1}(z)\; \chi^-_{0,1}(z)\; \chi^ +_{1,1}(z)\; \chi^-_{1,1}(z)\right)^t$, yielding the following expression for the Hamiltonian
\begin{widetext}
\begin{eqnarray}
H=\left( \begin{array}{cccc} \frac{-\hbar^2}{2m}\frac{d^2}{dz^2}+\epsilon+V_\alpha(z) & 0&0& r(z)e^{-2i\theta(z)}
\\ 0 & \frac{-\hbar^2}{2m}\frac{d^2}{dz^2}+\epsilon+V_\alpha(z) & r(z)e^{-2i\theta(z)}&0
\\ 0&+r(z)e^{2i\theta(z)}& \frac{-\hbar^2}{2m}\frac{d^2}{dz^2}+\lambda+V_\alpha(z)&\\ r(z)e^{2i\theta(z)}&0&0& \frac{-\hbar^2}{2m}\frac{d^2}{dz^2}+\lambda+V_\alpha(z)  \end{array} \right),\label{eq4}
\end{eqnarray}
\end{widetext}

where the SOC contribution, $V_\alpha(z)=-\frac{m{\alpha}(z)^2}{2\hbar^2}$, appears as an effective attractive quantum well potential, as also shown in Ref.~\citenum{Sanchez}. Its width is determined by $L$ and the depth, by the Rashba coupling strength (proportional to the perpendicular electric field $E_x$). 
Note that the eigenvalue problem involving the Hamiltonian of Eq.~(\ref{eq4}), leads to equations coupled in pairs with the same spin in the $\sigma_y$-basis, given by 
\begin{widetext}
\begin{eqnarray}
\frac{-\hbar^2}{2m}\frac{d^2\chi^\pm_{0,1}(z)}{dz^2}+[\epsilon-E-V_\alpha(z)]\chi^\pm_{0,1}(z)+r(z)e^{- 2i\theta(z)}\chi^\mp_{1,1}(z)=0,\label{eqc1}\\
\frac{-\hbar^2}{2m}\frac{d^2\chi^\pm_{1,1}(z)}{dz^2}+[\lambda-E-V_\alpha(z)]\chi^\pm_{1,1}(z)+r(z)e^{2i\theta(z)}\chi^\mp_{0,1}(z)=0\label{eqc2}
\end{eqnarray}
\end{widetext}

\begin{figure}[htb]
\includegraphics[scale=1.,angle=00]{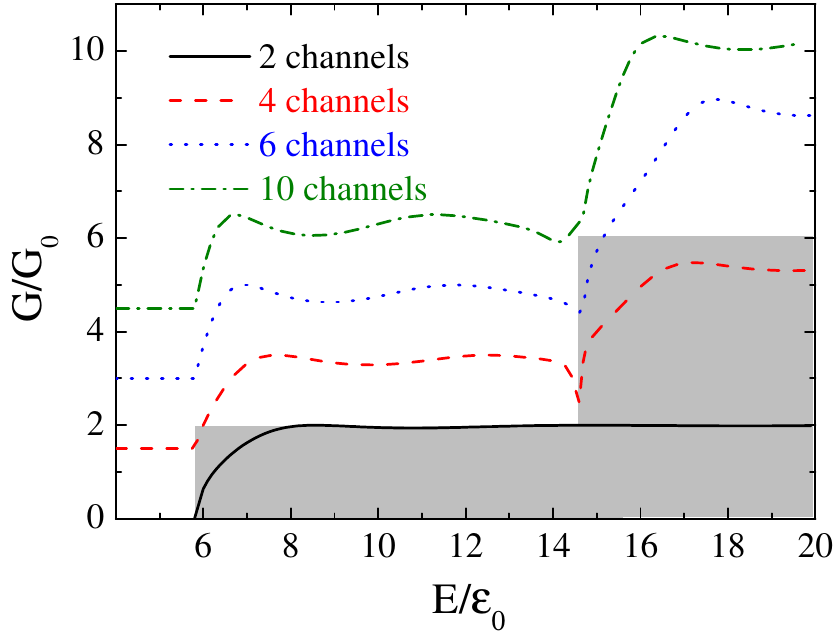}\caption{Normalized conductance as a function of the normalized Fermi energy for 2, 4, 6, and 10 channels, considering $L=1.4r_0$, $\alpha=3\alpha_0$, and null magnetic field.}\label{channels}
\end{figure}

\begin{figure}[htb]
\includegraphics[scale=1,angle=00]{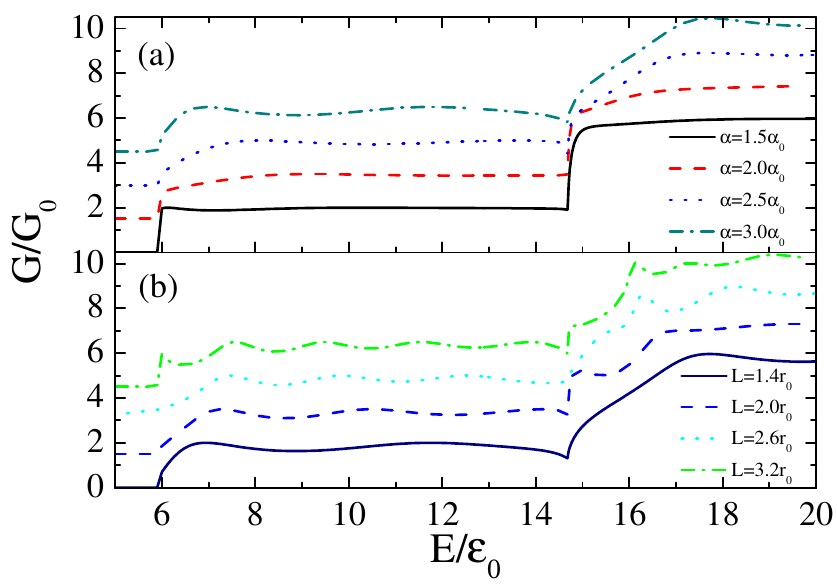}\caption{(a) Normalized conductance as a function of the normalized Fermi energy for 6 channels, considering $L=1.4r_0$, null magnetic field, and $\alpha=$1.5, 2, 2.5, and 3.0$\alpha_0$. (b) Normalized conductance as a function of the normalized Fermi energy for 6 channels, considering $L=$1.4, 2.0, 2.6, and 3.2 $r_0$, null magnetic field, and $\alpha=$ 3.0$\alpha_0$.}\label{alphas}
\end{figure}

\begin{figure}[htb]
\includegraphics[scale=1,angle=00]{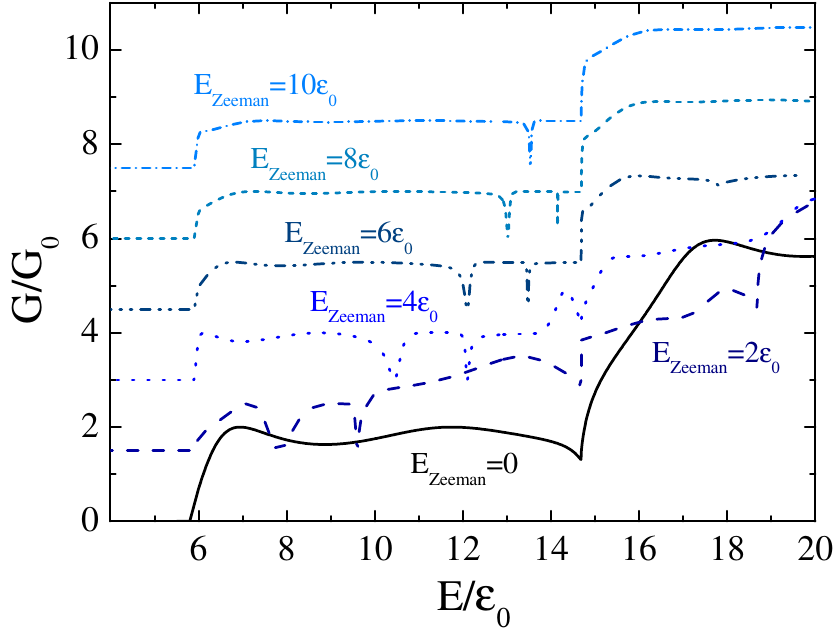}\caption{Normalized conductance as a function of the normalized Fermi energy for 6 channels, considering $L=1.4r_0$, $\alpha_R=$3.0, and different Zeeman energy $E_{Zeeman}$=0, 2, 4, 6, 8, and 10.}\label{zemman}
\end{figure}

Thus, by the mere presence of the Rashba SOC, the necessary conditions to observe the resonant reflection~\cite{Gurvitz} are satisfied, which are the existence of an attractive potential $V_\alpha(z)$ and the coupling between different channels ruled by the term $r(z)e^{\pm2i\theta(z)}$ as can be observed in Eqs.~(5) and (6). It is paramount to assert that a strict 1D transport model considering the Rashba SOC between different spins, will not lead to the requested conditions just described above because the lack of the attractive potential. In Fig.~\ref{channels}, we plot the numerical solution for the normalized conductance as a function of the total energy varying the number of scattering channels considering ${\alpha}=3\alpha_0$ and $L=1.4r_0$, where $\varepsilon_0=\hbar^2/2m^*r_0^2$ and $\alpha_0=\varepsilon_0r_0$ are the natural energy and Rashba constant scales. The total energy $E$ represents the applied bias, which in turn is equivalent to the variation of the gate-voltage applied perpendicularly to the NW~\cite{Rainis}. The numerical results were obtained by noting that the total Hamiltonian commutes with the momentum in the $z-$ direction $p_z$, therefore the scattering wave-functions can be expanded in the $\exp{(ik_zz)}$ basis. By using the expansion of the wave-function in the $\exp{(ik_zz)}$ basis and by matching the wave-functions at the interfaces $z=\pm L/2$, we get a system of linear equations. The solution of this system of linear equations yields all the coefficients of the expansion in the $\exp{(ik_zz)}$ basis plus all the transmission and reflection coefficients. The conductance is evaluated through the Landauer-Buttiker formula $G=G_0Tr\left[tt^\dagger \right]$, where $t$ is the matrix composed of the transmission elements $t_{i,j}$, where $i$ and $j$ denote the transmitted and the incident channels, respectively.
 Considering only two channels (similar to strict 1D transport model), the conductance is almost constant for $E> 8\varepsilon_0$ and no dip is observed as expected. However, when four scattering channels are taken into account, a strong reflection occurs, leading to a dip in the conductance, as can be observed in the plot of normalized conductance shown in Fig.~\ref{channels} for $E\approx14.5\varepsilon_0$. This phenomenon occurs because the intersubband Rashba SOC term is equivalent to an attractive potential in the region $|z|\leq L/2$ as can be seen from Eqs.(\ref{eq1},\ref{eq2}). Such a negative potential can support a quasibound state within this region and this state interacts with the continuum states causing the strong reflection due to interference effects. It is interesting to mention that for total energy less than $E\approx 14.67\epsilon_0$, only the quantum numbers $(0,1,\pm)$ admit scattering modes, whereas the other quantum numbers (channels) correspond to  evanescent modes. Therefore, the influence of evanescent modes in the transport properties is very important, as can be seen when we compare the conductance evaluated considering two and four channels. When more than four channels are considered, the reflection region broadens and suffers a redshift, but the main feature of the re-entrant behaviour is preserved. Hereafter, we will fix the number of channels to six, because the influence of high energy channels would exponentially decay if incoherent transport effects were considered~\cite{winkler}. Moreover, among the six first channels the states $(-1,1,\pm)$ are present, which are degenerated in eigenenergy with the third and fourth channels with quantum numbers $(-1,1,\pm)$. In Fig.~\ref{alphas} (a), we plot the normalized conductance using the same parameters of Fig.~\ref{channels}, except for the Rashba constant, which is varied as follows: ${\alpha}=1.5\alpha_0$ (solid curve), ${\alpha}=2.0\alpha_0$ (dotted curve), ${\alpha}=2.5\alpha_0$ (short dashed curve), and ${\alpha}=3.0\alpha_0$ (long dashed curve).  By decreasing the Rashba coupling constant (see Fig.\ref{alphas} (a)), the dip in the conductance can also be observed, but it becomes narrower and eventually shallower. The normalized conductance for different lengths of the region where the SOC takes place is shown in Fig.~\ref{alphas} (b), considering the following parameters: $L=$1.4$r_0$ (solid curve), $L=$2.0$r_0$ (dashed curve), $L=$2.6$r_0$ (dotted curve), and $L=$3.2$r_0$ (dashed-dotted curve), null magnetic field, and $\alpha=$ 3.0$\alpha_0$. More oscillations in the conductance appear when the length L is increased (see Fig.\ref{alphas} (b)), but the dip in the conductance for $E\approx14.5\varepsilon_0$ is barely affected. The increasing of the oscillations are due to Fabry-Perot interference~\cite{Sun,Rainis}, which depends on the length of the scattering region.

 In the experiment described in Ref.~\citenum{Heedt}, it was also observed that the dip in the conductance appears for zero and for moderate values of the magnetic field, but disappears for high values. To probe such effects in our theoretical model, we consider a uniform magnetic field applied in the $x$-direction described by the Hamiltonian $H^{Zeeman}=E_{Zeeman}\sigma_x$, where $E_{Zeeman}$ is the Zeeman energy. The matrix elements for this Hamiltonian are given by $H^Z_{j,j^\prime}=E_Z\delta_{n,n^\prime}\delta_{m,m^\prime}\delta_{s,-s^\prime}$. Unlike the Rashba SOC, the magnetic field operates not only in the region with length $L$, but throughout the whole NW. When the magnetic field is included, there is a shift in the subband energy $\epsilon^\pm_j=\epsilon_{n,m}\pm E_{Zeeman}$, which would cause a shift in the total energy where the first step in the conductance would be observed. On the other hand, the electron source will also be affected by the uniform magnetic field in the same way; thereby, cancelling out this energy shift. In other words, we consider that the total energy where the first channel opens does not change as the magnetic field varies. In Fig.~\ref{zemman}, we plot the normalized conductance for different values of the Zeeman energy, $E_{Zeeman}$ ,as a function of the total energy. For $E_{Zeeman}> 0$, the injection of only one electron occurs between $\epsilon_1\leq E\leq \epsilon_1+2E_{Zeeman}$ and the injection of two electrons occurs after $E>\epsilon_1+2E_{Zeeman}$. For $E_{Zeeman}=2\epsilon_0$, the dip at $E\approx14.5\varepsilon_0$ still appears, but other dips occur $E\approx7.8\varepsilon_0$ and $E\approx9.6\varepsilon_0$, which are related to interference processes that occur due to the attractive potential and the coupling between different transport channels. For $E_{Zeeman}=4\varepsilon_0$ and $E_{Zeeman}=6\varepsilon_0$, three dips appear in a similar way as the case $E_{Zeeman}=2\varepsilon_0$. For $E_{Zeeman}>6\varepsilon_0$, the dip at $E\approx14.5\varepsilon_0$ disappears and other dips become very narrow. Only one dip in the conductance is observed for $E_{Zeeman}=10\varepsilon_0$. We can thus ascribe the vanishing of the re-entrant behaviour for high magnetic fields observed in Ref.~\citenum{Heedt} to the combination of the dips narrowing produced by increasing the magnetic field plus incoherent transport processes. The incoherent transport smears out the re-entrant behaviour~\cite{Rainis} and for high values of the magnetic field it would not allow the appearance of very narrow regions of reflection.
 
 In conclusion, we presented a theoretical model capable of explaining the re-entrant behaviour of the conductance in the absence of the external magnetic field and that can be sustained also for high field values. The dip in the conductance appears because of the resonant reflection, which is predicted to occur in quasi-one-dimensional systems if two conditions are met: the existence of an attractive potential and the coupling between different scattering channels, even though the coupling occurs between a scattering- and an evanescent-mode. Both conditions coexist when considering the NW experiences the Rashba SOC in a limited region of length $L$. The importance of the limited regions is related to the existence of the effective attractive potential that occurs when Rashba SOC takes place within this region, which agrees with the experimental setup described in Ref.~\citenum{Heedt}.

The authors are grateful for financial support by the Brazilian Agencies FAPESP, CNPq and CAPES. L. K. C. acknowledges the support of FAPESP (grant No 2019/09624-3) for supporting this research. The data that supports the findings of this study are available within the article.

\bibliography{ref}

\begin{thebibliography}{19}%
\makeatletter
\providecommand \@ifxundefined [1]{%
 \@ifx{#1\undefined}
}%
\providecommand \@ifnum [1]{%
 \ifnum #1\expandafter \@firstoftwo
 \else \expandafter \@secondoftwo
 \fi
}%
\providecommand \@ifx [1]{%
 \ifx #1\expandafter \@firstoftwo
 \else \expandafter \@secondoftwo
 \fi
}%
\providecommand \natexlab [1]{#1}%
\providecommand \enquote  [1]{``#1''}%
\providecommand \bibnamefont  [1]{#1}%
\providecommand \bibfnamefont [1]{#1}%
\providecommand \citenamefont [1]{#1}%
\providecommand \href@noop [0]{\@secondoftwo}%
\providecommand \href [0]{\begingroup \@sanitize@url \@href}%
\providecommand \@href[1]{\@@startlink{#1}\@@href}%
\providecommand \@@href[1]{\endgroup#1\@@endlink}%
\providecommand \@sanitize@url [0]{\catcode `\\12\catcode `\$12\catcode
  `\&12\catcode `\#12\catcode `\^12\catcode `\_12\catcode `\%12\relax}%
\providecommand \@@startlink[1]{}%
\providecommand \@@endlink[0]{}%
\providecommand \url  [0]{\begingroup\@sanitize@url \@url }%
\providecommand \@url [1]{\endgroup\@href {#1}{\urlprefix }}%
\providecommand \urlprefix  [0]{URL }%
\providecommand \Eprint [0]{\href }%
\providecommand \doibase [0]{http://dx.doi.org/}%
\providecommand \selectlanguage [0]{\@gobble}%
\providecommand \bibinfo  [0]{\@secondoftwo}%
\providecommand \bibfield  [0]{\@secondoftwo}%
\providecommand \translation [1]{[#1]}%
\providecommand \BibitemOpen [0]{}%
\providecommand \bibitemStop [0]{}%
\providecommand \bibitemNoStop [0]{.\EOS\space}%
\providecommand \EOS [0]{\spacefactor3000\relax}%
\providecommand \BibitemShut  [1]{\csname bibitem#1\endcsname}%
\let\auto@bib@innerbib\@empty
\bibitem [{\citenamefont {Heedt}\ \emph {et~al.}(2017)\citenamefont {Heedt},
  \citenamefont {Ziani}, \citenamefont {Crepin}, \citenamefont {Prost},
  \citenamefont {Trellenkamp}, \citenamefont {Schubert}, \citenamefont
  {Gruetzmacher}, \citenamefont {Trauzettel},\ and\ \citenamefont
  {Schaepers}}]{Heedt}%
  \BibitemOpen
  \bibfield  {author} {\bibinfo {author} {\bibfnamefont {S.}~\bibnamefont
  {Heedt}}, \bibinfo {author} {\bibfnamefont {N.~T.}\ \bibnamefont {Ziani}},
  \bibinfo {author} {\bibfnamefont {F.}~\bibnamefont {Crepin}}, \bibinfo
  {author} {\bibfnamefont {W.}~\bibnamefont {Prost}}, \bibinfo {author}
  {\bibfnamefont {S.}~\bibnamefont {Trellenkamp}}, \bibinfo {author}
  {\bibfnamefont {J.}~\bibnamefont {Schubert}}, \bibinfo {author}
  {\bibfnamefont {D.}~\bibnamefont {Gruetzmacher}}, \bibinfo {author}
  {\bibfnamefont {B.}~\bibnamefont {Trauzettel}}, \ and\ \bibinfo {author}
  {\bibfnamefont {T.}~\bibnamefont {Schaepers}},\ }\href {\doibase
  10.1038/NPHYS4070} {\bibfield  {journal} {\bibinfo  {journal} {NATURE
  PHYSICS}\ }\textbf {\bibinfo {volume} {13}},\ \bibinfo {pages} {563}
  (\bibinfo {year} {2017})}\BibitemShut {NoStop}%
\bibitem [{\citenamefont {Nayak}\ \emph {et~al.}(2008)\citenamefont {Nayak},
  \citenamefont {Simon}, \citenamefont {Stern}, \citenamefont {Freedman},\ and\
  \citenamefont {Das~Sarma}}]{Nayak}%
  \BibitemOpen
  \bibfield  {author} {\bibinfo {author} {\bibfnamefont {C.}~\bibnamefont
  {Nayak}}, \bibinfo {author} {\bibfnamefont {S.~H.}\ \bibnamefont {Simon}},
  \bibinfo {author} {\bibfnamefont {A.}~\bibnamefont {Stern}}, \bibinfo
  {author} {\bibfnamefont {M.}~\bibnamefont {Freedman}}, \ and\ \bibinfo
  {author} {\bibfnamefont {S.}~\bibnamefont {Das~Sarma}},\ }\href {\doibase
  10.1103/RevModPhys.80.1083} {\bibfield  {journal} {\bibinfo  {journal}
  {REVIEWS OF MODERN PHYSICS}\ }\textbf {\bibinfo {volume} {80}},\ \bibinfo
  {pages} {1083} (\bibinfo {year} {2008})}\BibitemShut {NoStop}%
\bibitem [{\citenamefont {Alicea}\ \emph {et~al.}(2011)\citenamefont {Alicea},
  \citenamefont {Oreg}, \citenamefont {Refael}, \citenamefont {von Oppen},\
  and\ \citenamefont {Fisher}}]{Alicea}%
  \BibitemOpen
  \bibfield  {author} {\bibinfo {author} {\bibfnamefont {J.}~\bibnamefont
  {Alicea}}, \bibinfo {author} {\bibfnamefont {Y.}~\bibnamefont {Oreg}},
  \bibinfo {author} {\bibfnamefont {G.}~\bibnamefont {Refael}}, \bibinfo
  {author} {\bibfnamefont {F.}~\bibnamefont {von Oppen}}, \ and\ \bibinfo
  {author} {\bibfnamefont {M.~P.~A.}\ \bibnamefont {Fisher}},\ }\href {\doibase
  10.1038/NPHYS1915} {\bibfield  {journal} {\bibinfo  {journal} {NATURE
  PHYSICS}\ }\textbf {\bibinfo {volume} {7}},\ \bibinfo {pages} {412} (\bibinfo
  {year} {2011})}\BibitemShut {NoStop}%
\bibitem [{\citenamefont {Oreg}\ \emph {et~al.}(2010)\citenamefont {Oreg},
  \citenamefont {Refael},\ and\ \citenamefont {von Oppen}}]{Oreg}%
  \BibitemOpen
  \bibfield  {author} {\bibinfo {author} {\bibfnamefont {Y.}~\bibnamefont
  {Oreg}}, \bibinfo {author} {\bibfnamefont {G.}~\bibnamefont {Refael}}, \ and\
  \bibinfo {author} {\bibfnamefont {F.}~\bibnamefont {von Oppen}},\ }\href
  {\doibase 10.1103/PhysRevLett.105.177002} {\bibfield  {journal} {\bibinfo
  {journal} {PHYSICAL REVIEW LETTERS}\ }\textbf {\bibinfo {volume} {105}}
  (\bibinfo {year} {2010}),\ 10.1103/PhysRevLett.105.177002}\BibitemShut
  {NoStop}%
\bibitem [{\citenamefont {Lutchyn}\ \emph {et~al.}(2010)\citenamefont
  {Lutchyn}, \citenamefont {Sau},\ and\ \citenamefont {Das~Sarma}}]{Lutchyn}%
  \BibitemOpen
  \bibfield  {author} {\bibinfo {author} {\bibfnamefont {R.~M.}\ \bibnamefont
  {Lutchyn}}, \bibinfo {author} {\bibfnamefont {J.~D.}\ \bibnamefont {Sau}}, \
  and\ \bibinfo {author} {\bibfnamefont {S.}~\bibnamefont {Das~Sarma}},\ }\href
  {\doibase 10.1103/PhysRevLett.105.077001} {\bibfield  {journal} {\bibinfo
  {journal} {PHYSICAL REVIEW LETTERS}\ }\textbf {\bibinfo {volume} {105}}
  (\bibinfo {year} {2010}),\ 10.1103/PhysRevLett.105.077001}\BibitemShut
  {NoStop}%
\bibitem [{\citenamefont {Kammhuber}\ \emph {et~al.}(2017)\citenamefont
  {Kammhuber}, \citenamefont {Cassidy}, \citenamefont {Pei}, \citenamefont
  {Nowak}, \citenamefont {Vuik}, \citenamefont {Gul}, \citenamefont {Car},
  \citenamefont {Plissard}, \citenamefont {Bakkers}, \citenamefont {Wimmer},\
  and\ \citenamefont {Kouwenhoven}}]{Kammhuber}%
  \BibitemOpen
  \bibfield  {author} {\bibinfo {author} {\bibfnamefont {J.}~\bibnamefont
  {Kammhuber}}, \bibinfo {author} {\bibfnamefont {M.~C.}\ \bibnamefont
  {Cassidy}}, \bibinfo {author} {\bibfnamefont {F.}~\bibnamefont {Pei}},
  \bibinfo {author} {\bibfnamefont {M.~P.}\ \bibnamefont {Nowak}}, \bibinfo
  {author} {\bibfnamefont {A.}~\bibnamefont {Vuik}}, \bibinfo {author}
  {\bibfnamefont {O.}~\bibnamefont {Gul}}, \bibinfo {author} {\bibfnamefont
  {D.}~\bibnamefont {Car}}, \bibinfo {author} {\bibfnamefont {S.~R.}\
  \bibnamefont {Plissard}}, \bibinfo {author} {\bibfnamefont {E.~P. A.~M.}\
  \bibnamefont {Bakkers}}, \bibinfo {author} {\bibfnamefont {M.}~\bibnamefont
  {Wimmer}}, \ and\ \bibinfo {author} {\bibfnamefont {L.~P.}\ \bibnamefont
  {Kouwenhoven}},\ }\href {\doibase 10.1038/s41467-017-00315-y} {\bibfield
  {journal} {\bibinfo  {journal} {NATURE COMMUNICATIONS}\ }\textbf {\bibinfo
  {volume} {8}} (\bibinfo {year} {2017}),\
  10.1038/s41467-017-00315-y}\BibitemShut {NoStop}%
\bibitem [{\citenamefont {van Weperen}\ \emph {et~al.}(2013)\citenamefont {van
  Weperen}, \citenamefont {Plissard}, \citenamefont {Bakkers}, \citenamefont
  {Frolov},\ and\ \citenamefont {Kouwenhoven}}]{Weperen}%
  \BibitemOpen
  \bibfield  {author} {\bibinfo {author} {\bibfnamefont {I.}~\bibnamefont {van
  Weperen}}, \bibinfo {author} {\bibfnamefont {S.~R.}\ \bibnamefont
  {Plissard}}, \bibinfo {author} {\bibfnamefont {E.~P. A.~M.}\ \bibnamefont
  {Bakkers}}, \bibinfo {author} {\bibfnamefont {S.~M.}\ \bibnamefont {Frolov}},
  \ and\ \bibinfo {author} {\bibfnamefont {L.~P.}\ \bibnamefont
  {Kouwenhoven}},\ }\href {\doibase 10.1021/nl3035256} {\bibfield  {journal}
  {\bibinfo  {journal} {NANO LETTERS}\ }\textbf {\bibinfo {volume} {13}},\
  \bibinfo {pages} {387} (\bibinfo {year} {2013})}\BibitemShut {NoStop}%
\bibitem [{\citenamefont {Sun}\ \emph {et~al.}(2018)\citenamefont {Sun},
  \citenamefont {Deacon}, \citenamefont {Wang}, \citenamefont {Yao},
  \citenamefont {Lieber},\ and\ \citenamefont {Ishibashi}}]{Sun}%
  \BibitemOpen
  \bibfield  {author} {\bibinfo {author} {\bibfnamefont {J.}~\bibnamefont
  {Sun}}, \bibinfo {author} {\bibfnamefont {R.~S.}\ \bibnamefont {Deacon}},
  \bibinfo {author} {\bibfnamefont {R.}~\bibnamefont {Wang}}, \bibinfo {author}
  {\bibfnamefont {J.}~\bibnamefont {Yao}}, \bibinfo {author} {\bibfnamefont
  {C.~M.}\ \bibnamefont {Lieber}}, \ and\ \bibinfo {author} {\bibfnamefont
  {K.}~\bibnamefont {Ishibashi}},\ }\href {\doibase
  10.1021/acs.nanolett.8b01799} {\bibfield  {journal} {\bibinfo  {journal}
  {NANO LETTERS}\ }\textbf {\bibinfo {volume} {18}},\ \bibinfo {pages} {6144}
  (\bibinfo {year} {2018})}\BibitemShut {NoStop}%
\bibitem [{\citenamefont {Streda}\ and\ \citenamefont {Seba}(2003)}]{Streda}%
  \BibitemOpen
  \bibfield  {author} {\bibinfo {author} {\bibfnamefont {P.}~\bibnamefont
  {Streda}}\ and\ \bibinfo {author} {\bibfnamefont {P.}~\bibnamefont {Seba}},\
  }\href {\doibase 10.1103/PhysRevLett.90.256601} {\bibfield  {journal}
  {\bibinfo  {journal} {PHYSICAL REVIEW LETTERS}\ }\textbf {\bibinfo {volume}
  {90}} (\bibinfo {year} {2003}),\ 10.1103/PhysRevLett.90.256601}\BibitemShut
  {NoStop}%
\bibitem [{\citenamefont {Khrapai}\ and\ \citenamefont
  {Nagaev}(2018)}]{Khrapai}%
  \BibitemOpen
  \bibfield  {author} {\bibinfo {author} {\bibfnamefont {V.~S.}\ \bibnamefont
  {Khrapai}}\ and\ \bibinfo {author} {\bibfnamefont {K.~E.}\ \bibnamefont
  {Nagaev}},\ }\href {\doibase 10.1103/PhysRevB.98.121401} {\bibfield
  {journal} {\bibinfo  {journal} {PHYSICAL REVIEW B}\ }\textbf {\bibinfo
  {volume} {98}} (\bibinfo {year} {2018}),\
  10.1103/PhysRevB.98.121401}\BibitemShut {NoStop}%
\bibitem [{\citenamefont {Saldana}\ \emph {et~al.}(2018)\citenamefont
  {Saldana}, \citenamefont {Niquet}, \citenamefont {Cleuziou}, \citenamefont
  {Lee}, \citenamefont {Car}, \citenamefont {Plissard}, \citenamefont
  {Bakkers},\ and\ \citenamefont {De~Franceschi}}]{Saldana}%
  \BibitemOpen
  \bibfield  {author} {\bibinfo {author} {\bibfnamefont {J.~C.~E.}\
  \bibnamefont {Saldana}}, \bibinfo {author} {\bibfnamefont {Y.-M.}\
  \bibnamefont {Niquet}}, \bibinfo {author} {\bibfnamefont {J.-P.}\
  \bibnamefont {Cleuziou}}, \bibinfo {author} {\bibfnamefont {E.~J.~H.}\
  \bibnamefont {Lee}}, \bibinfo {author} {\bibfnamefont {D.}~\bibnamefont
  {Car}}, \bibinfo {author} {\bibfnamefont {S.~R.}\ \bibnamefont {Plissard}},
  \bibinfo {author} {\bibfnamefont {E.~P. A.~M.}\ \bibnamefont {Bakkers}}, \
  and\ \bibinfo {author} {\bibfnamefont {S.}~\bibnamefont {De~Franceschi}},\
  }\href {\doibase 10.1021/acs.nanolett.7b03854} {\bibfield  {journal}
  {\bibinfo  {journal} {NANO LETTERS}\ }\textbf {\bibinfo {volume} {18}},\
  \bibinfo {pages} {2282} (\bibinfo {year} {2018})}\BibitemShut {NoStop}%
\bibitem [{\citenamefont {Bardarson}\ \emph {et~al.}(2004)\citenamefont
  {Bardarson}, \citenamefont {Magnusdottir}, \citenamefont {Gudmundsdottir},
  \citenamefont {Tang}, \citenamefont {Manolescu},\ and\ \citenamefont
  {Gudmundsson}}]{Bardarson}%
  \BibitemOpen
  \bibfield  {author} {\bibinfo {author} {\bibfnamefont {J.}~\bibnamefont
  {Bardarson}}, \bibinfo {author} {\bibfnamefont {I.}~\bibnamefont
  {Magnusdottir}}, \bibinfo {author} {\bibfnamefont {G.}~\bibnamefont
  {Gudmundsdottir}}, \bibinfo {author} {\bibfnamefont {C.}~\bibnamefont
  {Tang}}, \bibinfo {author} {\bibfnamefont {A.}~\bibnamefont {Manolescu}}, \
  and\ \bibinfo {author} {\bibfnamefont {V.}~\bibnamefont {Gudmundsson}},\
  }\href {\doibase 10.1103/PhysRevB.70.245308} {\bibfield  {journal} {\bibinfo
  {journal} {PHYSICAL REVIEW B}\ }\textbf {\bibinfo {volume} {70}} (\bibinfo
  {year} {2004}),\ 10.1103/PhysRevB.70.245308}\BibitemShut {NoStop}%
\bibitem [{\citenamefont {Pershin}\ \emph {et~al.}(2004)\citenamefont
  {Pershin}, \citenamefont {Nesteroff},\ and\ \citenamefont
  {Privman}}]{Pershin}%
  \BibitemOpen
  \bibfield  {author} {\bibinfo {author} {\bibfnamefont {Y.}~\bibnamefont
  {Pershin}}, \bibinfo {author} {\bibfnamefont {J.}~\bibnamefont {Nesteroff}},
  \ and\ \bibinfo {author} {\bibfnamefont {V.}~\bibnamefont {Privman}},\ }\href
  {\doibase 10.1103/PhysRevB.69.121306} {\bibfield  {journal} {\bibinfo
  {journal} {PHYSICAL REVIEW B}\ }\textbf {\bibinfo {volume} {69}} (\bibinfo
  {year} {2004}),\ 10.1103/PhysRevB.69.121306}\BibitemShut {NoStop}%
\bibitem [{\citenamefont {Tang}\ \emph {et~al.}(2017)\citenamefont {Tang},
  \citenamefont {Yu}, \citenamefont {Abdullah},\ and\ \citenamefont
  {Gudmundsson}}]{Tang}%
  \BibitemOpen
  \bibfield  {author} {\bibinfo {author} {\bibfnamefont {C.-S.}\ \bibnamefont
  {Tang}}, \bibinfo {author} {\bibfnamefont {Y.-H.}\ \bibnamefont {Yu}},
  \bibinfo {author} {\bibfnamefont {N.~R.}\ \bibnamefont {Abdullah}}, \ and\
  \bibinfo {author} {\bibfnamefont {V.}~\bibnamefont {Gudmundsson}},\ }\href
  {\doibase 10.1016/j.physleta.2017.10.012} {\bibfield  {journal} {\bibinfo
  {journal} {PHYSICS LETTERS A}\ }\textbf {\bibinfo {volume} {381}},\ \bibinfo
  {pages} {3960} (\bibinfo {year} {2017})}\BibitemShut {NoStop}%
\bibitem [{\citenamefont {Rainis}\ and\ \citenamefont {Loss}(2014)}]{Rainis}%
  \BibitemOpen
  \bibfield  {author} {\bibinfo {author} {\bibfnamefont {D.}~\bibnamefont
  {Rainis}}\ and\ \bibinfo {author} {\bibfnamefont {D.}~\bibnamefont {Loss}},\
  }\href {\doibase 10.1103/PhysRevB.90.235415} {\bibfield  {journal} {\bibinfo
  {journal} {PHYSICAL REVIEW B}\ }\textbf {\bibinfo {volume} {90}} (\bibinfo
  {year} {2014}),\ 10.1103/PhysRevB.90.235415}\BibitemShut {NoStop}%
\bibitem [{\citenamefont {Sanchez}\ and\ \citenamefont
  {Serra}(2006)}]{Sanchez}%
  \BibitemOpen
  \bibfield  {author} {\bibinfo {author} {\bibfnamefont {D.}~\bibnamefont
  {Sanchez}}\ and\ \bibinfo {author} {\bibfnamefont {L.}~\bibnamefont
  {Serra}},\ }\href {\doibase 10.1103/PhysRevB.74.153313} {\bibfield  {journal}
  {\bibinfo  {journal} {PHYSICAL REVIEW B}\ }\textbf {\bibinfo {volume} {74}}
  (\bibinfo {year} {2006}),\ 10.1103/PhysRevB.74.153313}\BibitemShut {NoStop}%
\bibitem [{\citenamefont {GURVITZ}\ and\ \citenamefont
  {LEVINSON}(1993)}]{Gurvitz}%
  \BibitemOpen
  \bibfield  {author} {\bibinfo {author} {\bibfnamefont {S.}~\bibnamefont
  {GURVITZ}}\ and\ \bibinfo {author} {\bibfnamefont {Y.}~\bibnamefont
  {LEVINSON}},\ }\href {\doibase 10.1103/PhysRevB.47.10578} {\bibfield
  {journal} {\bibinfo  {journal} {PHYSICAL REVIEW B}\ }\textbf {\bibinfo
  {volume} {47}},\ \bibinfo {pages} {10578} (\bibinfo {year}
  {1993})}\BibitemShut {NoStop}%
\bibitem [{\citenamefont {Villegas-Lelovsky}\ \emph {et~al.}(2009)\citenamefont
  {Villegas-Lelovsky}, \citenamefont {Trallero-Giner}, \citenamefont {Rebello
  Sousa~Dias}, \citenamefont {Lopez-Richard},\ and\ \citenamefont
  {Marques}}]{leo}%
  \BibitemOpen
  \bibfield  {author} {\bibinfo {author} {\bibfnamefont {L.}~\bibnamefont
  {Villegas-Lelovsky}}, \bibinfo {author} {\bibfnamefont {C.}~\bibnamefont
  {Trallero-Giner}}, \bibinfo {author} {\bibfnamefont {M.}~\bibnamefont
  {Rebello Sousa~Dias}}, \bibinfo {author} {\bibfnamefont {V.}~\bibnamefont
  {Lopez-Richard}}, \ and\ \bibinfo {author} {\bibfnamefont {G.~E.}\
  \bibnamefont {Marques}},\ }\href {\doibase 10.1103/PhysRevB.79.155306}
  {\bibfield  {journal} {\bibinfo  {journal} {PHYSICAL REVIEW B}\ }\textbf
  {\bibinfo {volume} {79}} (\bibinfo {year} {2009}),\
  10.1103/PhysRevB.79.155306}\BibitemShut {NoStop}%
\bibitem [{\citenamefont {Winkler}(2003)}]{winkler}%
  \BibitemOpen
  \bibfield  {author} {\bibinfo {author} {\bibfnamefont {R.}~\bibnamefont
  {Winkler}},\ }\href@noop {} {\emph {\bibinfo {title} {Spin-Orbit Coupling
  Effects in Two-Dimensional Electron and Hole Systems}}}\ (\bibinfo
  {publisher} {Springer Tracts in Modern Physics Springer},\ \bibinfo {address}
  {Berlin},\ \bibinfo {year} {2003})\BibitemShut {NoStop}%
\end{thebibliography}%
\end{document}